\documentclass[twocolumn]{aastex631}

\usepackage{amsmath}
\usepackage{amssymb}

\usepackage[T1]{fontenc}

\newcommand{\ketju}{{KETJU}}

\newcommand{\gadget}{\mbox{GADGET-3}}

\newcommand{\pn}[1]{#1PN}
\newcommand{\gal}[1]{g#1}
\newcommand{\bh}[2][]{#2$ #1$}

\renewcommand{\edit}[2]{#2}

\begin{document}
\title{
Signatures of the Many Supermassive Black Hole Mergers in a Cosmologically Forming Massive Early-Type Galaxy
}

\author[0000-0001-5721-9335]{Matias Mannerkoski}
\affiliation{
Department of Physics,
Gustaf H\"allstr\"omin katu 2, FI-00014, University of Helsinki, Finland
}

\author[0000-0001-8741-8263]{Peter H. Johansson}
\affiliation{
Department of Physics,
Gustaf H\"allstr\"omin katu 2, FI-00014, University of Helsinki, Finland
}

\author[0000-0001-8789-2571]{Antti Rantala}
\affiliation{
Max-Planck-Institut f\"ur Astrophysik, Karl-Schwarzchild-Str 1,
D-85748 Garching, Germany
}

\author[0000-0002-7314-2558]{Thorsten Naab}
\affiliation{
Max-Planck-Institut f\"ur Astrophysik, Karl-Schwarzchild-Str 1,
D-85748 Garching, Germany
}

\author[0000-0001-7075-6098]{Shihong Liao}
\affiliation{
Department of Physics,
Gustaf H\"allstr\"omin katu 2, FI-00014, University of Helsinki, Finland
}

\author[0000-0003-1807-6321]{Alexander Rawlings}
\affiliation{
Department of Physics,
Gustaf H\"allstr\"omin katu 2, FI-00014, University of Helsinki, Finland
}

\correspondingauthor{Matias Mannerkoski}
\email{matias.mannerkoski@helsinki.fi}

\begin{abstract}
We model here the merger histories of the supermassive black hole (SMBH)
population in the late stages of
a cosmological simulation of a $\sim 2 \times 10^{13} M_\sun$  galaxy group.
The gravitational dynamics around the several tens of SMBHs ($M_{\bullet} > 7.5\times 10^7 M_\sun$)
hosted by the galaxies in the group is computed at high accuracy
using regularized integration with the \ketju{} code.
The 11 SMBHs which form binaries and a hierarchical triplet eventually merge after
hardening through dynamical friction, stellar scattering, and gravitational wave (GW) emission.
The binaries form at eccentricities of  $e \sim 0.3 \textnormal{--} 0.9$,
with one system evolving to a very high eccentricity of $e = 0.998$,
and merge on timescales of a few tens to several hundred megayears.
During the simulation, the merger-induced GW recoil kicks eject one SMBH remnant from the central host galaxy.
This temporarily drives the galaxy off the $M_{\bullet} \textnormal{--} \sigma_{\star}$ relation,
however the galaxy returns to the relation due to subsequent galaxy mergers, which bring in new SMBHs.
This showcases a possible mechanism contributing to the observed scatter of the \edit2{$M_{\bullet} \textnormal{--} \sigma_{\star}$} relation.
Finally, we show that Pulsar Timing Arrays and LISA would be able to detect parts of the GW signals from
the SMBH mergers that occur during the $\sim 4\,\mathrm{Gyr}$ time span simulated with \ketju{}.

\end{abstract}

\section{Introduction}
The masses of the central supermassive black holes (SMBHs) in massive galaxies
are tightly correlated with the structural properties of their host galaxies,
which is manifested in the observed $M_{\bullet} \textnormal{--} \sigma_{\star}$ relation
\citep[e.g.][]{kormendy2013}.
\edit2{Massive early-type galaxies are believed to have assembled 
through a two-stage process, in which the early assembly is dominated by rapid in situ star formation in gas-rich systems, whereas 
the later growth below redshifts of $z\lesssim 2\textnormal{--}3$ is dominated by relatively gas-poor minor mergers \citep[e.g.][]{2012ApJ...754..115J,2017ARA&A..55...59N}}. 
During galaxy mergers,
SMBHs merge through a three-stage process \citep{begelman1980},
driven first by dynamical friction until a binary forms,
then by three-body scattering between the SMBH binary and
individual stars \citep{1980AJ.....85.1281H}, and finally at subparsec scales
by gravitational wave (GW) emission \citep{1964PhRv..136.1224P}. 

The GW emission from a binary is in general asymmetric due to the different SMBH
masses and spins.
This produces a recoil kick,
typically giving the merged SMBH a velocity $\lesssim 500\,\mathrm{km\,s^{-1}}$,
but reaching $\sim 4000\,\mathrm{km\,s^{-1}}$ for suitable spins \citep[e.g.][]{2007ApJ...659L...5C}.
Large kick velocities can significantly displace the merged SMBH or even eject
it from the galaxy. 
This has been suggested to give rise to the observed offset active galactic nuclei
\citep[e.g.][]{2015ApJ...806..219C}, 
drive the formation of large galactic cores \citep{2021MNRAS.502.4794N},
and contribute to the scatter in the
$M_\bullet \textnormal{--} \sigma_\star$ relation \citep{2007ApJ...663L...5V,2011MNRAS.412.2154B}.

The GWs emitted during the final phases of an SMBH merger could be detectable by ongoing 
Pulsar Timing Array (PTA) projects, which primarily target the stochastic superposition
of GWs from the expected large number of SMBH binaries \citep[e.g.][]{2020ApJ...905L..34A} but
may also detect individual loud sources.
The PTAs are most sensitive 
to GWs in the nanohertz frequency range and therefore target
very  massive SMBHs ${M_{\bullet} \gtrsim 10^{8} M_\sun}$,
with orbital periods of around a few years \citep[e.g.][]{2017MNRAS.471.4508K}.
The Laser Interferometer Space Antenna (LISA) \citep{2017arXiv170200786A} is a
planned space-based detector that will mainly target the GWs emitted during the
inspiral and merger of slightly lower-mass SMBHs of $M_{\bullet} \sim 10^6 \textnormal{--}10^7 M_\sun$ up to very high redshifts.
However, the final parts of the signals from low-redshift mergers of SMBH binaries
with masses around $M_{\bullet} \sim 10^8 M_\sun$ are
also expected to be detectable with LISA \citep{2019MNRAS.483.3108K}. 

\edit2{
The small-scale dynamics of SMBHs in merging galaxies have been studied extensively in isolated collisionless
galaxy mergers
\citep[e.g.][]{2009ApJ...695..455B,2011ApJ...732...89K,2018ApJ...864..113R,2021MNRAS.502.4794N}, however, 
to date only very few high-resolution cosmological zoom-in simulations with detailed SMBH dynamics have been 
carried out \citep{2016ApJ...828...73K,2021ApJ...912L..20M}.
Instead, studies of merging SMBH binaries in a cosmological context have typically made use of
semi-analytic methods \citep[e.g.][]{2017MNRAS.471.4508K,2019MNRAS.486.4044B,2022MNRAS.509.3488I},
which make several assumptions about the unresolved binary orbits, some of which may not be fully motivated.
}

In this \edit2{paper} we study the dynamics of SMBHs and their GW signals in a 
cosmological zoom-in simulation of a group-sized
halo hosting dozens of SMBHs with masses $M_{\bullet} \gtrsim 10^8 M_\sun$.
The simulation is run using our 
updated \ketju{} code \citep{2017ApJ...840...53R,2018ApJ...864..113R,2021ApJ...912L..20M}, which is 
able to resolve the dynamics of merging SMBH down to tens of Schwarzschild radii.
We now also include a model for GW recoil kicks, which allows us to study the displacement 
and ejection of SMBHs from their host galaxies. 

\edit2{This paper is structured as follows. In Section \ref{Sec:sims} we give a brief overview of the \ketju{} code and
describe our simulation setup. In Section \ref{sec:results} we first describe the general properties of the simulated SMBH 
mergers, highlighting also the structural and kinematic properties of their host galaxies. This is followed by a discussion of the 
evolution of the SMBHs on the {$M_{\bullet} \textnormal{--} \sigma_{\star}$} plane. We round off this section with a calculation 
of the detectability of our simulated GW signals.
In Section \ref{Sec:discuss} the validity and implications of our results are discussed, and finally, in Section \ref{Sec:conclusions}
 we present our conclusions.}

\section{Numerical Simulations}
\label{Sec:sims}

\begin{figure*}
\centering
\includegraphics{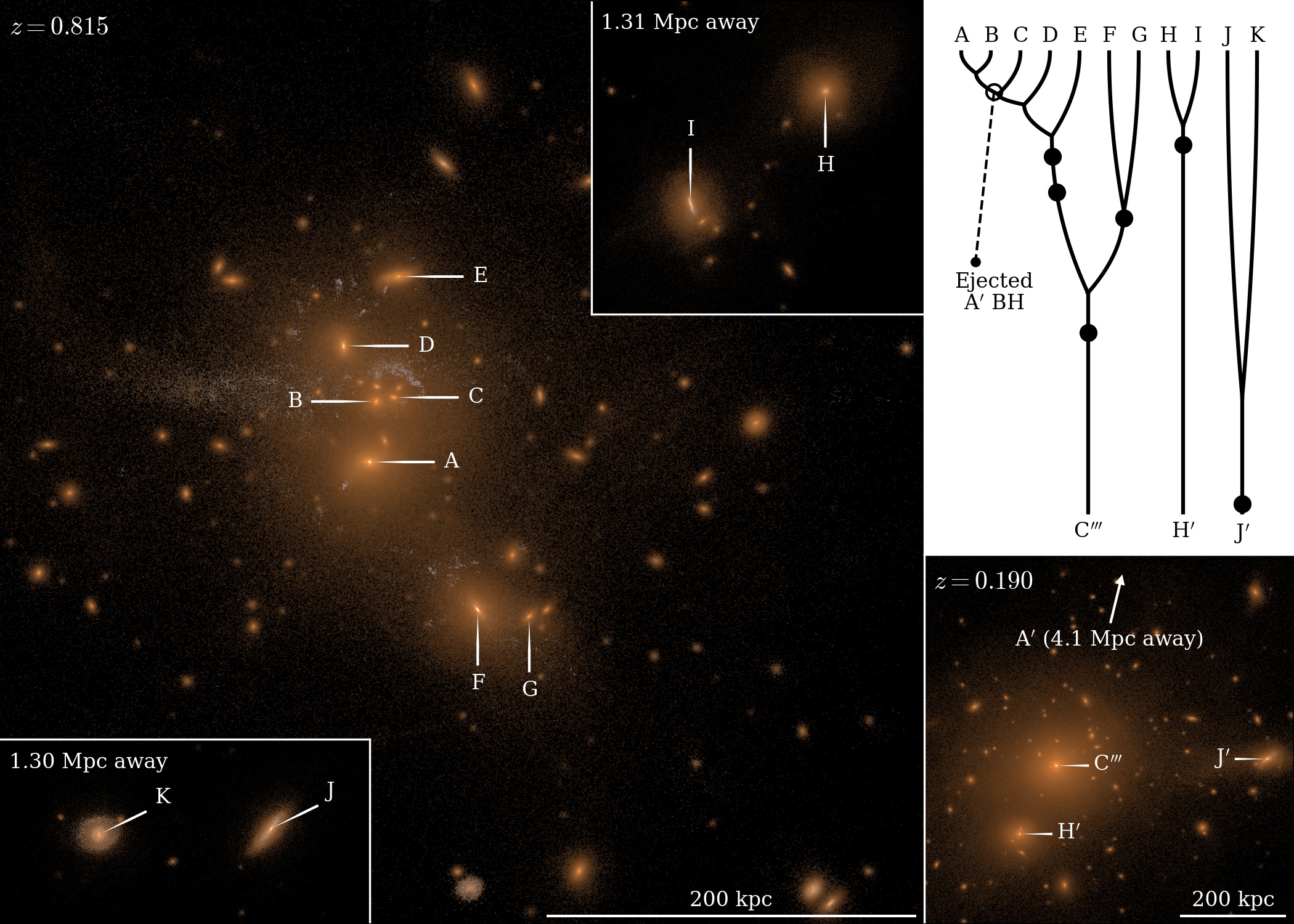}
\caption{
Left: 
The galaxies and SMBHs of interest in the initial state of the \ketju{} run.
The main panel shows the central group of galaxies,
with the two more distant pairs of galaxies shown in the corners at the same spatial scale.
The images are generated from BVR-luminosities accounting only for stellar emission.
Top right:
A schematic merger tree of the galaxies and SMBHs with time
progressing from top to bottom.
The lines follow the galaxy mergers, while the circles indicate SMBH binary mergers.
Bottom right:
The final state of the simulation with the remaining SMBHs labeled.
}
\label{fig:overview}
\end{figure*}

\subsection{The \ketju{} Code}

The \ketju{} code extends
the widely-used \gadget{} code \citep{2005MNRAS.364.1105S} by replacing the 
standard leapfrog integration of SMBHs and their surrounding stellar particles
with the high-accuracy regularized MSTAR integrator 
\citep{2020MNRAS.492.4131R} in a small
region around each SMBH.
\edit2{
All SMBH-SMBH and SMBH-stellar particle interactions in the regularized regions
are computed without gravitational softening, while the gravitational interactions between stellar particles
are softened to avoid energy errors when particles enter and exit the regions.
The center-of-mass motion of the regularized regions is integrated using the same leapfrog integrator
that is used for all the other particles, and the small perturbing tidal
forces acting on the regions are also included through leapfrog kicks instead of
the computationally more costly but also slightly more accurate perturber particle
method used in earlier \ketju{} versions
\citep{2017ApJ...840...53R,2018ApJ...864..113R}.
}

Post-Newtonian (PN) corrections are included in interactions between SMBHs to account for relativistic effects
such as GW emission.
In addition to binary PN terms,
we now also include the leading order \pn{1} corrections of a general $N$-body system,
which contain terms involving up to three bodies that may affect the long-term
evolution of triplet SMBH systems \citep{2014PhRvD..89d4043W,2020PhRvD.102f4033L}. 
This is done using the expressions from \cite{1985PhRvD..31.1815T}
for the \pn{1} and spin terms, whereas higher-order corrections valid for BH binary systems
up to \pn{3.5} order are adopted from Eq. (203) of \cite{2014LRR....17....2B}.

\edit2{In addition}, we now include a model for the mass, spin, and recoil velocity of SMBH
merger remnants based on the numerical relativity fitting functions from \cite{2015PhRvD..92b4022Z}.
The model uses as inputs the spins and orbital orientation of the BH binary when
it is merged at a separation of $12 G (M_1 + M_2) /c^2$,
where $M_{1,2}$ are the component masses.
This gives remnant properties that account for the particular spin directions
and orbital orientation of the SMBHs determined by their prior evolution,
but which are still only approximate due to the inherent limitations of the fit functions. 

We use the same hydrodynamics, star formation, and feedback models as in \cite{2021ApJ...912L..20M}.
\edit2{The hydrodynamics of gas is modeled using the SPHGal smoothed particle hydrodynamics (SPH) 
implementation \citep{2014MNRAS.443.1173H}, which employs a pressure-entropy formulation together with artificial conduction and 
viscosity utilizing a Wendland $C^{4}$-kernel smoothed over 100 neighbors.}
Our models include metal-dependent cooling that tracks 11 individual elements \edit2{and
stochastic star formation with a critical hydrogen number density threshold of $n_{\rm H}=0.1 \, \mathrm{cm}^{-3}$} 
\citep{2005MNRAS.364..552S,2006MNRAS.371.1125S,2013MNRAS.434.3142A}. \edit2{In addition we include
models for feedback from supernovae and massive stars, as well as the production of metals
through stellar chemical evolution \citep{2013MNRAS.434.3142A,2017MNRAS.468..751E}.}

\edit2{Galaxies with dark matter halo masses of  $M_{\rm DM}=10^{10} h^{-1} M_{\odot}$ are 
seeded with black holes with masses of $M_{\bullet}=10^{5} h^{-1} M_{\odot}$, which then grow through gas
accretion and merging \citep{2007MNRAS.380..877S}}. The accretion onto SMBHs is modeled using a simple Bondi--Hoyle--Lyttleton 
prescription \edit2{with an additional dimensionless multiplier of $\alpha=25$ to account for the limited spatial resolution \citep{2009ApJ...690..802J}.} 
\edit2{The maximum accretion rate is capped at the Eddington limit and assuming a fixed radiative efficiency 
of $\epsilon_{r}=0.1$ a total of 0.5\% of the rest mass energy of the accreted gas is coupled to the 
surrounding gas as thermal energy \citep{2005MNRAS.361..776S}.}
This model successfully produces SMBHs in agreement with the observed scaling relations,
but does not correctly model the accretion onto tight binaries or the evolution of SMBH spin.
However, these shortcomings are not significant,
as the SMBH binaries in this study are \edit2{predominantly} found in gas-poor
environments and have \edit2{thus low gas accretion rates}.

\subsection{Initial Conditions and Simulations}

We run a cosmological zoom-in simulation starting at a redshift \edit2{of} $z=50$,
targeting a much larger volume than in \cite{2021ApJ...912L..20M}.
The high-resolution region is centered on a dark matter (DM) halo with a
virial mass of $M_{200} \sim 2.5\times 10^{13} M_\sun$, covering an initial comoving size of 
$(10 h^{-1}\,\mathrm{Mpc})^3$ and containing $410^3$ of both gas and DM particles
with masses of $m_\mathrm{gas}={3 \times 10^5 M_\sun}$ and  $m_\mathrm{DM}={1.6 \times 10^6 M_\sun}$. 
The initial conditions are generated with the  MUSIC \citep{2011MNRAS.415.2101H}
software package using the Planck 2018 cosmology \citep{2020A&A...641A...6P} with
a Hubble parameter of $h=0.674$.

We first evolve this volume with standard \gadget{} including SMBH seeding and
repositioning until $z=0.815$, at which point the simulation contains 
tens of galaxies hosting SMBHs with masses $>7.5 \times 10^{7} M_\sun$.
We take this as the lower mass limit of SMBHs to be modeled with
\ketju{}, since a high enough SMBH to stellar particle mass ratio is required
to ensure accurate binary dynamics \edit2{\citep{2021ApJ...912L..20M}.}
For this initial run,
the gravitational softening lengths were set to $\epsilon_\mathrm{bar}={40 h^{-1} \,\mathrm{pc}}$
for stars and gas, and
$\epsilon_\mathrm{DM,high}={93 h^{-1}\,\mathrm{pc}}$ for high-resolution dark matter particles.

We continue the run from $z=0.815$ with \ketju{}.
For this run, we lower the softening
length of the stellar particles to $\epsilon_\mathrm{\star}={20 h^{-1} \,\mathrm{pc}}$
to allow using regularized regions with radii of ${60 h^{-1}\,\mathrm{pc}}$,
\edit2{resulting in a manageable level of a few thousand stellar particles in each regularized region.}

At this point we also add spins to the SMBH particles to model their merger recoil kicks.
The spin directions are generated from a uniform distribution
on the sphere, whereas the spin magnitudes use the distribution
from \cite{2010PhRvD..81h4023L},
resulting in dimensionless spins $\chi =  c J / G M^2 \sim 0.5\textnormal{--}0.9$,
where $J$ is the spin angular momentum.
\edit2{
Observations constrain the spins of SMBHs with masses above $10^8 M_\sun$ rather poorly,
but are consistent with their spins being in this range \citep{2019NatAs...3...41R}.
During the simulation, the dimensionless spin 
may decrease slightly as the gas accretion model does not yet evolve the spin angular momentum,
while the mass of the SMBHs may increase.}

\section{Results}
\label{sec:results}

\subsection{Galaxy and SMBH Mergers}

\begin{deluxetable*}{ccccccccc}
\tablecaption{
Properties of the SMBHs and Their Host Galaxies
\label{table:bh_properties}
}

\tablehead{
\colhead{BH label\tablenotemark{a}} 
& \colhead{$M_{\bullet}/10^8 M_\sun$\tablenotemark{b}}
& \colhead{$M_{\bullet}/m_\star$\tablenotemark{c}}
& \colhead{$\chi$\tablenotemark{d}}
& \colhead{$v_\mathrm{kick}/\mathrm{km\,s^{-1}}$\tablenotemark{e}}
& \colhead{$M_\star/10^{10} M_\sun$\tablenotemark{f}}
& \colhead{$M_\mathrm{gas}/M_\star$\tablenotemark{g}}
& \colhead{$R_{1/2}/\mathrm{kpc}$\tablenotemark{h}}
& \colhead{$N_\star(R_{1/2})/10^5$\tablenotemark{i}}
}

\startdata
\bh{A} & 12.2 & 4400 & 0.84 & & 29.3 & 0.000 & 1.7 &4.64
\\
\bh{B} &  6.7 & 2500 & 0.77 & & 10.1 & 0.007 & 1.2 &1.66
\\
\bh{C} &  2.6 &  900 & 0.63 & &  9.5 & 0.006 & 3.2 &1.62
\\
\bh{D} &  6.7 & 2400 & 0.74 & & 13.4 & 0.001 & 1.0 &1.92
\\
\bh{E} &  3.3 & 1200 & 0.48 & & 10.2 & 0.000 & 1.2 &1.65
\\
\bh{F} &  7.9 & 2800 & 0.79 & & 14.5 & 0.023 & 0.8 &1.89
\\
\bh{G} &  1.2 &  500 & 0.81 & &  3.3 & 0.067 & 0.7 &0.47
\\
\bh{H} &  3.3 & 1200 & 0.74 & &  8.0 & 0.017 & 1.5 &1.35
\\
\bh{I} &  3.3 & 1200 & 0.58 & &  8.7 & 0.057 & 1.4 &1.36
\\
\bh{J} &  2.2 &  800 & 0.62 & &  6.2 & 0.088 & 1.2 &0.89
\\
\bh{K} &  1.1 &  400 & 0.59 & &  4.9 & 0.182 & 1.4 &0.84
\\
\bh{A}\bh{B} $\to$ \bh[']{A} & 18.1 & 6700 & 0.67 & 2257 & & & & 
\\
\bh{H}\bh{I} $\to$ \bh[']{H} &  6.6 & 2400 & 0.62 &  137 & 12.5 & 0.000 & 2.1 &2.09
\\
\bh{C}\bh{D} $\to$ \bh[']{C} &  9.6 & 3600 & 0.53 &  431 & & & & 
\\
\bh[']{C}\bh{E} $\to$ \bh['']{C} & 13.0 & 4800 & 0.70 &  492 & & & & 
\\
\bh{F}\bh{G} $\to$ \bh[']{F} &  9.3 & 3400 & 0.56 &  511 & & & & 
\\
\bh['']{C}\bh[']{F} $\to$ \bh[''']{C} & 23.1 & 8500 & 0.69 &  537 & 52.4 & 0.000 & 2.5 &8.02
\\
\bh{J}\bh{K} $\to$ \bh[']{J} &  5.3 & 2000 & 0.64 &  195 & 10.9 & 0.011 & 1.3 &1.61
\enddata

\tablenotetext{a}{Component labels $\to$ remnant label for merger remnants.}
\tablenotetext{b}{SMBH mass.}
\tablenotetext{c}{\edit1{SMBH to stellar particle mass ratio.}}
\tablenotetext{d}{SMBH dimensionless spin parameter.}
\tablenotetext{e}{Merger recoil kick velocity.}
\tablenotetext{f}{\edit1{Stellar mass within $r< 15\,\mathrm{kpc}$.}}
\tablenotetext{g}{\edit1{Gas fraction within $r< 15\,\mathrm{kpc}$.}}
\tablenotetext{h}{\edit1{Projected stellar half-mass radius.}}
\tablenotetext{i}{\edit1{Number of stellar particles within $R_{1/2}$.}}
\end{deluxetable*}

\begin{figure*}
\centering
\includegraphics{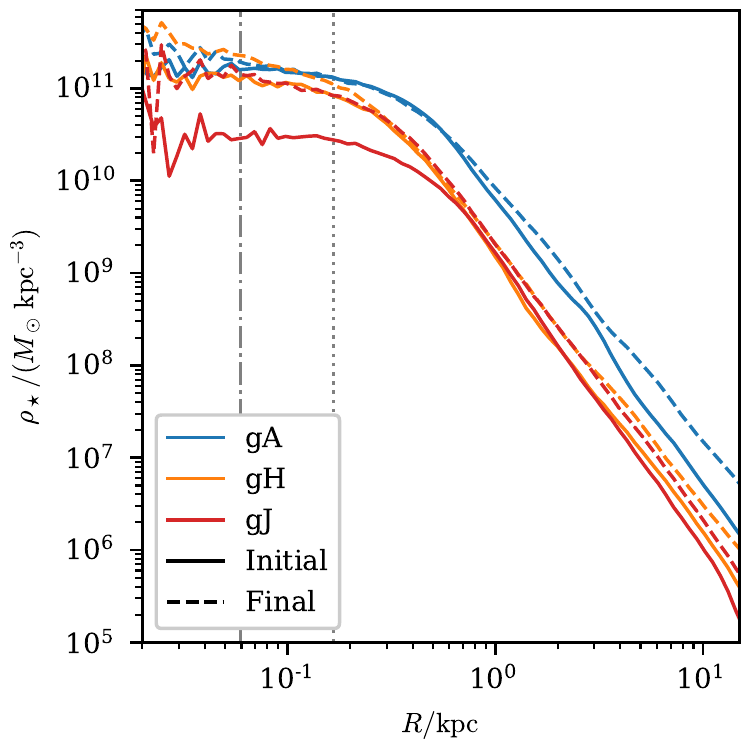}
\includegraphics{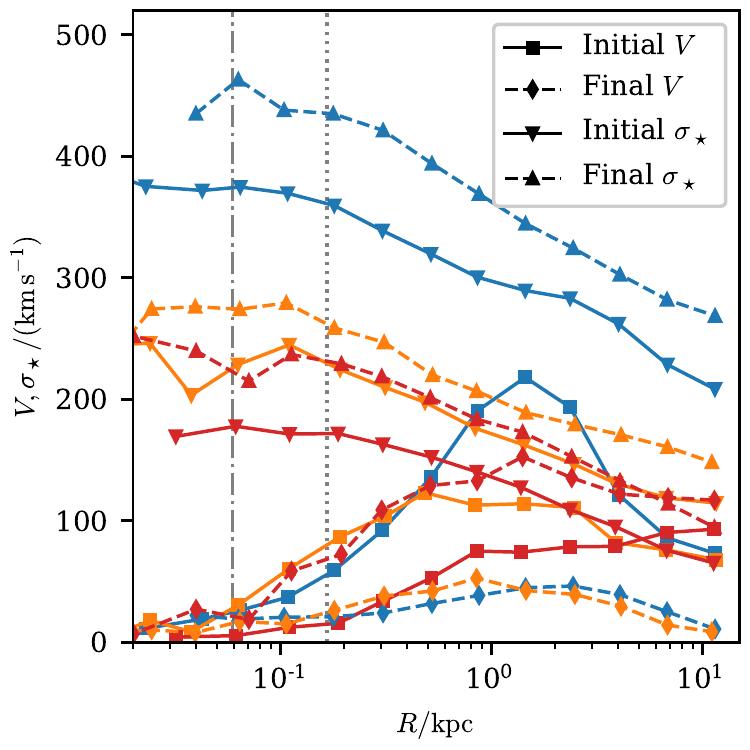}
\includegraphics{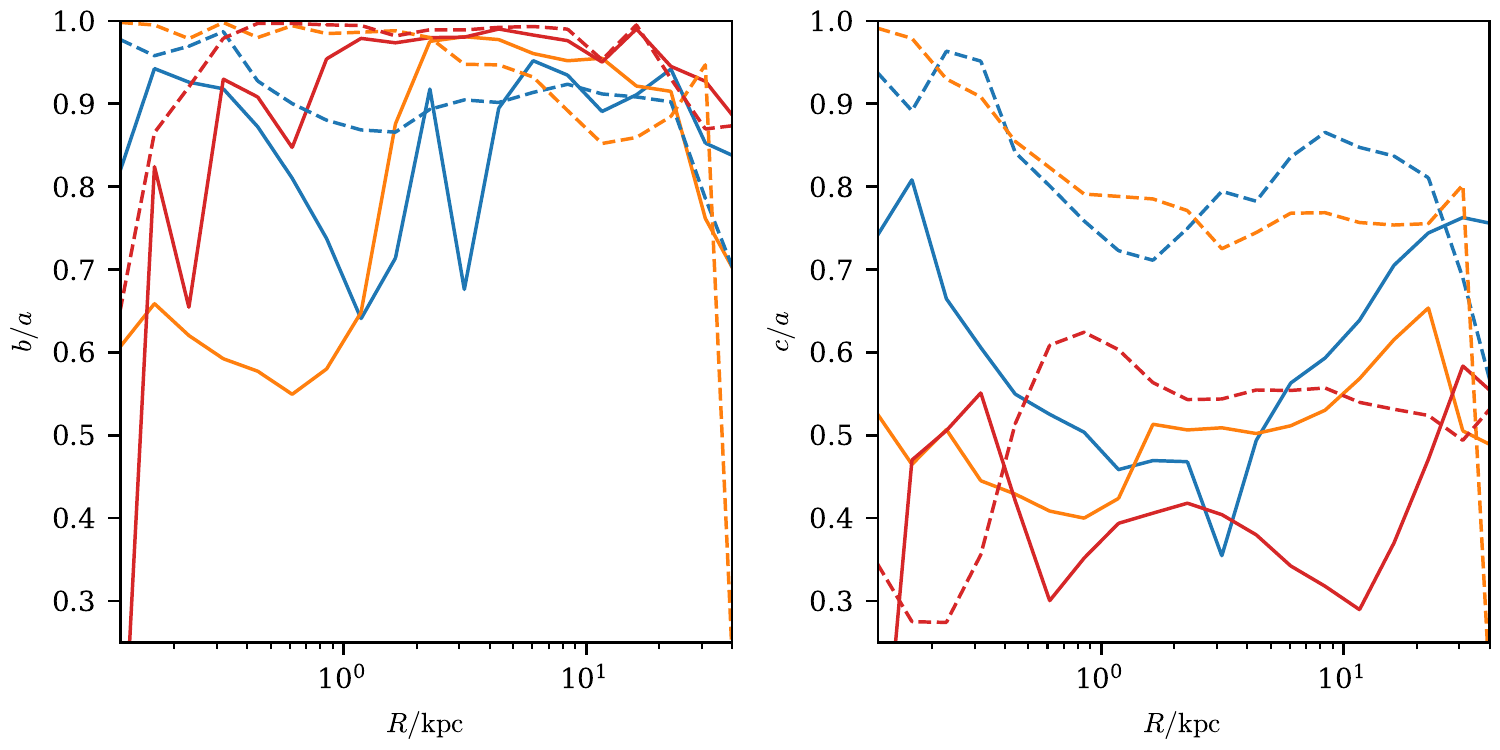}
\caption{
\edit1{
Top left:
Three-dimensional stellar density profiles in the initial and final states of the
\ketju{} run of the three galaxies \edit2{(gA, gH and gJ)} remaining in the final output.
The vertical dash-dotted and dotted lines show the $\epsilon_\mathrm{bar}=40 h^{-1}\,\mathrm{pc}$
softening length and the $2.8\epsilon_\mathrm{bar}$ softening kernel size used in the initial \gadget{} run.
Top right: Projected stellar rotational velocity $V$ and velocity dispersion
$\sigma_\star$ profiles of the same galaxies measured along the major rotation axis.
}
\edit2{
Bottom:
The axis ratio profiles of the galaxies computed using the S1 method of \cite{2011ApJS..197...30Z}.
}
}
\label{fig:galaxy_profiles}
\end{figure*}

\begin{figure*}
\includegraphics{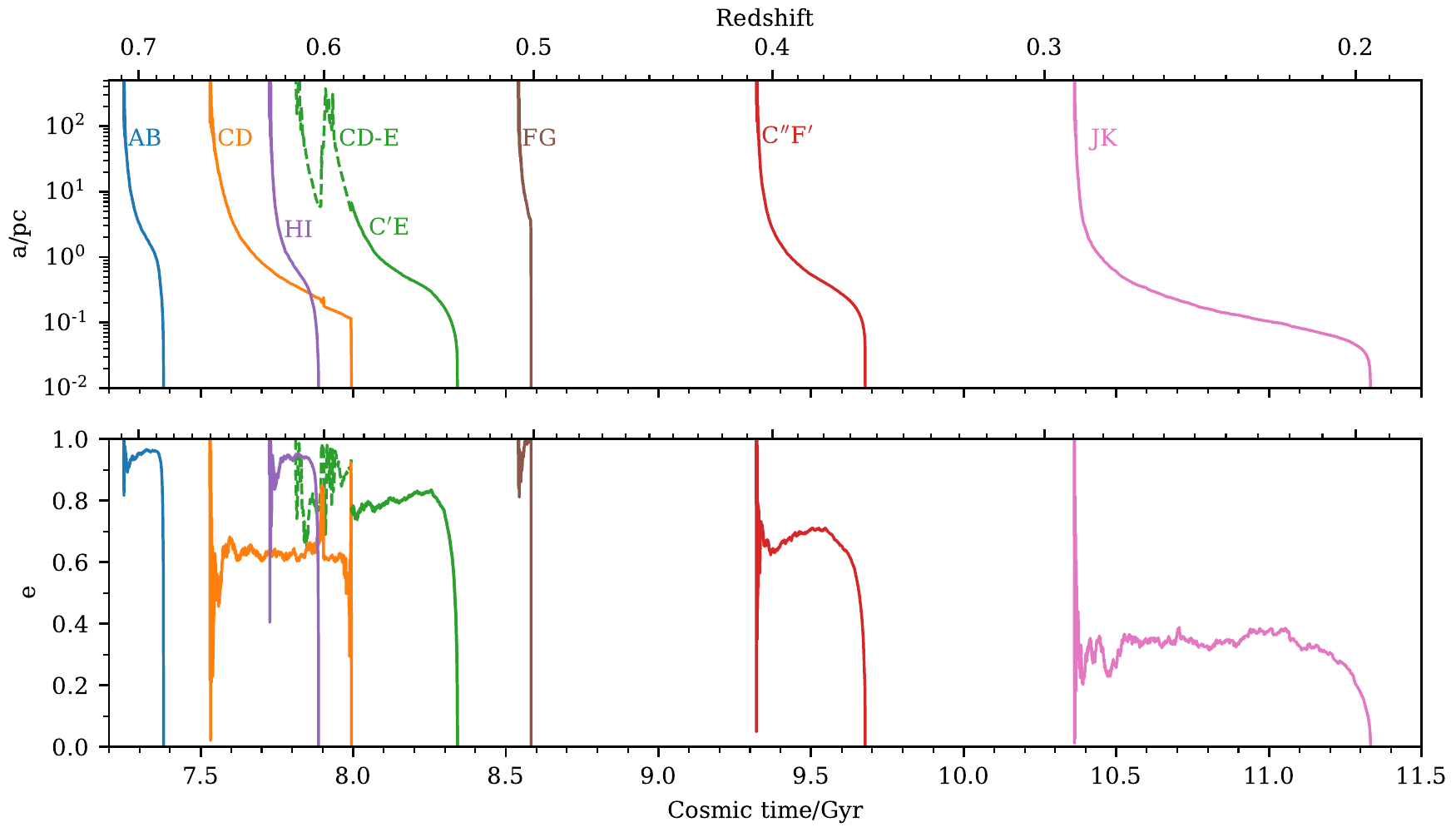}
\caption{
\edit2{The evolution of the} semi-major axis $a$ (top) and \edit2{the} eccentricity $e$ (bottom) of the \edit2{simulated} SMBH binaries.
The dashed line indicates the parameters of the outer orbit in the hierarchical
CD-E triplet.
}
\label{fig:binary_params}
\end{figure*}

At the start of the \ketju{} simulation there are a total of 11 galaxies hosting
massive SMBHs that are involved in mergers over the course of the simulation.
These galaxies are shown in the left panel of \autoref{fig:overview},
with the labels marking the locations of the SMBHs (e.g. A), with 
the host galaxies being referred to as e.g.\ \gal{A}.

Seven of the galaxies (\gal{A}--\gal{G}) are located in a central group that is collapsing
within a halo with a total virial mass of $M_{200} \approx 2\times 10^{13} M_\sun$
and a virial radius of $R_{200} \approx 420\,\mathrm{kpc}$.
In addition, there are two pairs of galaxies (\gal{H} and \gal{I}, \gal{K} and \gal{J})
on their initial orbits before merging. 
They are located within halos of 
$M_{200} \approx 2.5\times 10^{12} M_\sun$,
$R_{200} \approx 220\,\mathrm{kpc}$ for the \gal{H}-\gal{I} pair and
$M_{200} \approx 1.3\times 10^{12} M_\sun$,
$R_{200} \approx 170\,\mathrm{kpc}$ for the \gal{J}-\gal{K} pair.
\edit1{The properties of these galaxies and their SMBHs are listed in \autoref{table:bh_properties}.}
Finally, there are another 19 galaxies in the simulation volume, which host SMBHs that are \edit2{to begin with} or \edit2{later} grow to be
massive enough to be modeled with \ketju{}, as well as a large
number of lower-mass SMBHs.
However, they do not interact with the \edit2{more massive} SMBHs studied here. 

\edit2{
In \autoref{fig:galaxy_profiles} we show the density, velocity, and axis ratio
profiles of three selected galaxies (gA, gH, and gJ).
The galaxies are clearly triaxial at the start of the \ketju{} run,
and show moderately fast rotation with $V/\sigma_\star \sim 0.5$ in their outer parts.
The density profiles are flattened in the centers, resulting in galactic cores
with radii of a few hundred parsec, which are caused by the impact of gravitational softening.
However, in general the structural and kinematic properties of our simulated galaxies 
are in good agreement with the observed properties of $z=0.8$ galaxies \citep[e.g.][]{2018ApJ...858...60B},
although the half-mass radii of our simulated galaxies are toward the lower end of the observed range.
}

\edit2{
As the simulation progresses, the galaxies merge in the order shown in the 
top right panel of \autoref{fig:overview}.
The final state of the simulation at $z=0.19$ is shown in the lower right panel of \autoref{fig:overview}.
The simulation is stopped at this point as there are no more imminent galaxy mergers involving massive SMBHs.
The properties of the galaxies in this final output are also listed in \autoref{table:bh_properties},
and their density, velocity, and shape profiles are included in \autoref{fig:galaxy_profiles}.
The galaxies evolve as expected, with the sizes and velocity dispersions increasing,
while the rotational velocities and triaxialities mainly decrease.
The central flattening of the density profile is also somewhat reduced, due to the fact that the SMBHs
are resolved as non-softened point masses.
}

\edit2{
The galaxy mergers result in bound SMBH binaries that harden and finally merge
due to stellar interactions and GW emission.
The evolution of the PN-corrected orbital parameters
\citep{2004PhRvD..70j4011M,2019ApJ...887...35M}
of the binaries is shown in \autoref{fig:binary_params},
and the properties of the merger remnants are
listed in \autoref{table:bh_properties}.
}

Most of the SMBH binaries form at moderately
high eccentricities of $e = 0.6 \textnormal{--} 0.95$,
with \edit2{limited eccentricity evolution} during the hardening process.
These high eccentricities result in relatively short binary lifetimes
of ${\sim 200 \textnormal{--} 500 \,\mathrm{Myr}}$.
However, the \bh{F}\bh{G}-binary is an exception to this general trend, with a very high 
peak eccentricity of $e = 0.998$ resulting in an extremely rapid GW-driven 
merger in just a few tens of megayears.
\edit2{
The eccentricity growth occurs when the binary semi-major axis is still above $10\,\mathrm{pc}$,
and the mass ratio of the binary is also relatively large $(\sim 7:1)$, which suggests that 
resonant dynamical friction \citep{1996NewA....1..149R} might be at work
in addition to the eccentricity growth caused by simple stellar scattering \citep{1996NewA....1...35Q}.
}
Another exception \edit2{to the trend of moderately high eccentricities} is the low-eccentricity ($e\approx 0.35$) \bh{J}\bh{K}-binary,
which forms after a nearly circular orbit galaxy merger and takes almost a gigayear to merge.
The host galaxies \gal{J} and \gal{K} are also gas-rich, 
leading to significant gas accretion onto the SMBHs during the simulation,
\edit2{however this accretion does not markedly affect the binary evolution}.

Similarly to \cite{2021ApJ...912L..20M}, a SMBH triplet also occurs in this
simulation (\bh{C}\bh{D}-\bh{E} in \autoref{fig:binary_params}).
\edit2{
After being temporarily ejected to a wider orbit through strong gravitational interactions with
the \bh{C}\bh{D}-binary, SMBH \bh{E} settles into a hierarchical triplet configuration
around the inner binary.
}
However, contrary to our previous study, the outer orbital period is in this case shorter than the relativistic
precession period of the inner binary.
This results in von Zeipel--Lidov--Kozai type oscillations \citep{1962P&SS....9..719L} that
eventually excite the \bh{C}\bh{D}-binary eccentricity from $e\approx 0.55$ to a high value of
$e \approx 0.9$.
At the relatively small semi-major axis of $a\approx 0.12 \,\mathrm{pc}$ the 
increased eccentricity
is enough to cause a near instant GW-driven merger of the  \bh{C}\bh{D}-binary.

\subsection{SMBH Ejection and $M_\bullet$--$\sigma_\star$ Relation}

\begin{figure*}
\centering
\includegraphics{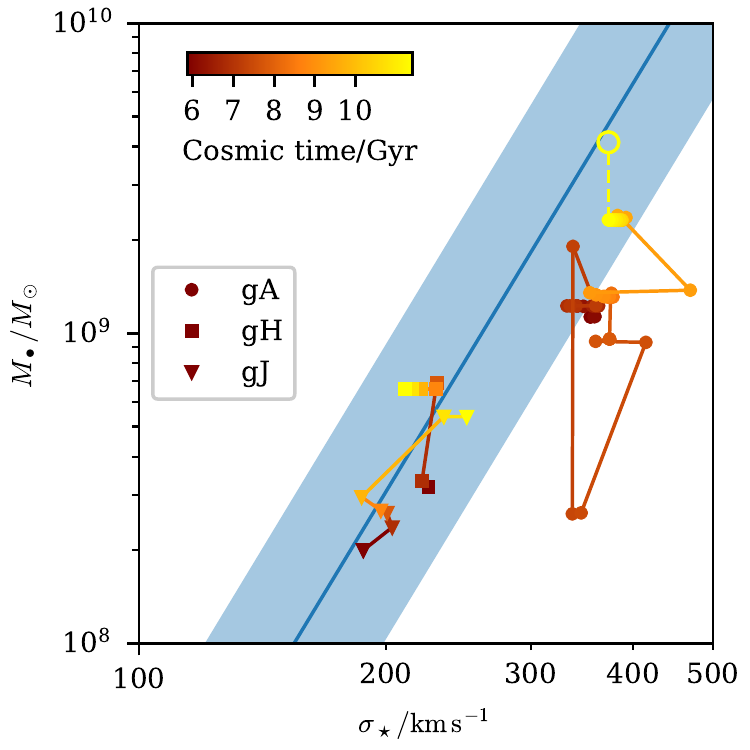}
\includegraphics{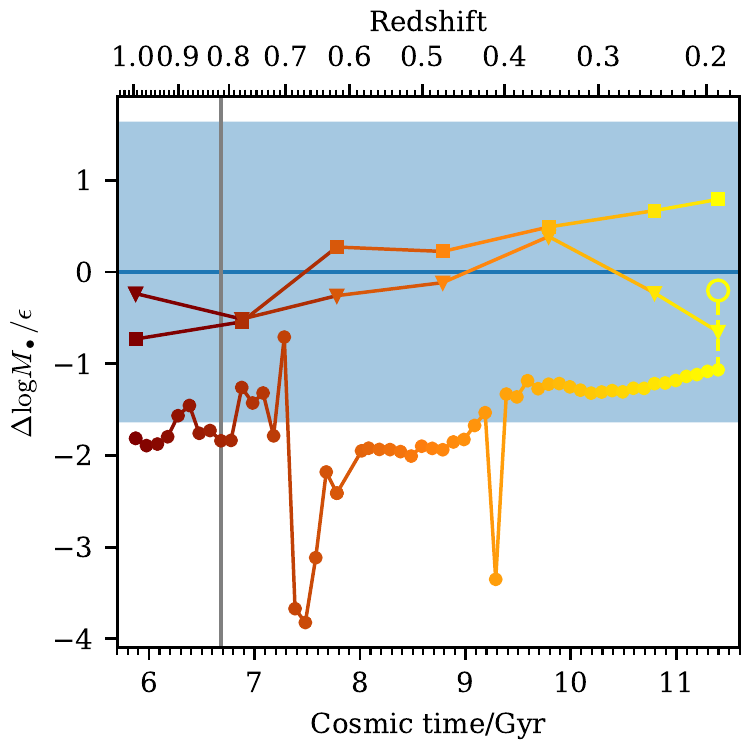}
\caption{
Left:
The time evolution of \edit2{the same galaxies (gA, gH and gJ) as in \autoref{fig:galaxy_profiles}}
on the $M_\bullet \textnormal{--} \sigma_\star$  plane. 
The line and shaded region show the observed relation (Eq. \ref{eq:bh-sigma})
and the $90\%$ prediction interval corresponding to the intrinsic scatter.
The open circle with a dashed line shows where the BH mass would lie 
without the ejection of \bh[']{A} from the gA-system.
Right:
The time evolution of the difference in mass to the relation in units of the
intrinsic scatter $\epsilon$ (see also \citealt{2009ApJ...707L.184J}).
The vertical \edit1{grey} line marks the start of the \ketju{} run.
}
\label{fig:bh_sigma}
\end{figure*}

The SMBH merger remnants \edit2{typically experience rather} modest GW recoil kicks 
\edit2{with velocities below $\lesssim 500\,\mathrm{km\,s^{-1}}$} (\autoref{table:bh_properties}).
Such kicks are not strong enough to displace the SMBHs significantly from the centers of
their host galaxies, and typically only result in oscillations of
$\sim 200\,\mathrm{pc}$
around the center of the galaxy that are dampened by dynamical
friction over a timescale of $\sim 10\,\mathrm{Myr}$.
The one exception is \bh[']{A},
which receives a kick of $2257\,\mathrm{km\,s^{-1}}$. 
This is above the galactic escape velocity $(v_\mathrm{esc}\approx 1950 \,\mathrm{km\,s^{-1}})$ and  \bh[']{A}
is thus ejected from its host galaxy, 
receding to a distance of $4.1 \, \mathrm{Mpc}$ by the end of the simulation.
While such large kick velocities are in general rare, we find that for this particular
SMBH spin configuration, the probability of 
a kick above $2000\,\mathrm{km\,s^{-1}}$ is about $20\%$.

The ejection of \bh[']{A} 
leads to an interesting evolution of its host galaxy \gal{A} on the 
$M_\bullet \textnormal{--} \sigma_\star$ plane, \edit2{as shown in} 
\autoref{fig:bh_sigma}. This figure also shows the evolution of 
galaxies \gal{H} and \gal{J} which undergo mergers without SMBH ejections.
The results from the simulation are compared to the observed relation by \citet{kormendy2013},
\begin{equation}
\label{eq:bh-sigma}
\log\left(\frac{M_\bullet}{10^9 M_\sun}\right) = -0.51 + 4.38 \log\left(\frac{\sigma_\star}{200\,\mathrm{km\,s^{-1}}}\right),
\end{equation}
which has an intrinsic scatter of $\epsilon=0.29$ in $\log{M_\bullet}$.

Initially \gal{A} lies within the range expected from the observed relation,
but after \bh[']{A} is ejected and \bh{C} enters the galaxy,
it becomes significantly offset from the relation at a cosmic time of $t=7.5 \ \rm Gyr$.
As galaxies \gal{D} and \gal{E} merge with \gal{A} they bring in their \edit2{central} SMBHs, resulting
in a partial recovery of the \edit2{expected} BH mass, with \gal{A} moving toward the observed relation.  
By the end of the simulation the offset falls within the 90\% region of the intrinsic scatter of the relation.
However, had \bh[']{A} not been ejected,  the galaxy and its
SMBH would lie even closer to the expected relation as is shown in \autoref{fig:bh_sigma} by the
open circle corresponding to a shift by the ejected SMBH mass.
In contrast, galaxies \gal{H} and \gal{J}  evolve following the observed relation after their respective mergers.

\subsection{Detectability of GW Signals with PTAs}

We can estimate whether the GW signal from our simulated SMBH binaries would be individually detectable with
PTAs by calculating the signal-to-noise ratio ($S/N$) including the effects of orbital
eccentricity using Eq. (65) from \cite{2015PhRvD..92f3010H}.
\edit2{
This method is similar to the semi-analytic GW spectrum calculation that was found to
work well in \cite{2019ApJ...887...35M}.
}
We use parameter values which resemble values found in
currently operating PTAs \cite[e.g.][]{2021ApJS..252....5A},
setting the 
number of pulsars to $N_p = 40$, the observation cadence to $\Delta t = 0.058\,\mathrm{yr}$
and the timing noise to $\sigma_\mathrm{rms} = 200 \,\mathrm{ns}$.
For the duration of observations we choose $T_\mathrm{obs} = 25\,\mathrm{yr}$, as this
increases the detectability of individual systems considerably compared to the
current operation time of $\sim 12\,\mathrm{yr}$.

The evolution of the $S/N$ values before the mergers of the
binaries are shown in the left panel of \autoref{fig:gw_signals}.
Following \cite{2016ApJ...817...70T},
binaries with $S/N$ above $\sim 5$ might be detectable as resolved sources, meaning 
that four out of seven \edit2{of our simulated} SMBH binaries would
be detectable starting from $\sim 10\,\mathrm{Myr}$ before the merger.
The other binaries would likely not be individually detectable,
but would produce a significant contribution to the GW background.

Unlike circular binaries, which only emit GWs with a frequency $f = 2 f_\mathrm{orb}$,
eccentric binaries emit a signal containing all harmonics $f = n f_\mathrm{orb}$ of
the orbital frequency.
This results in the gradual increase and saw-like oscillation of $S/N$,
as different harmonics pass through the fairly
sharp region of highest sensitivity of the PTA.
For the \bh{A}\bh{B} and \bh['']{C}\bh[']{F} binaries the signal is detectable
already during a period with significant eccentricity, so that $\sim 1\,\mathrm{Myr}$
prior to the merger the detected GW signal comes from higher harmonics that are not present for a circular binary.
In contrast the other binaries only become detectable when the orbits are close to
circular and the signal is dominated by the $n=2$ harmonic.

\subsection{Detectability of GW Signals with LISA}
The massive SMBH binaries in this study are not prime targets for the planned LISA
mission \citep{2017arXiv170200786A}, as most of their inspiral signal falls below
the target frequency sensitivity lower limit of $f=2\times10^{-5}\,\mathrm{Hz}$.
However, the GW signals from the merger and ringdown phases might still be detectable. In order to test this
 we calculate the spectra and sky averaged $S/N$ for our SMBH mergers
using the BOWIE code \citep{2019MNRAS.483.3108K} and its implementation of the
PhenomD waveform \citep{2016PhRvD..93d4006H,2016PhRvD..93d4007K}.

\begin{figure*}
\centering
\includegraphics{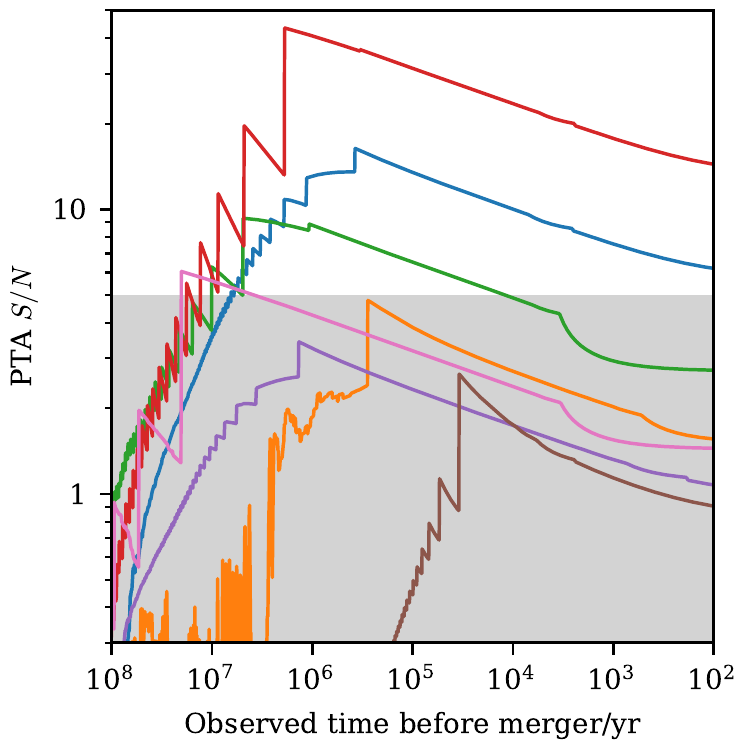}
\includegraphics{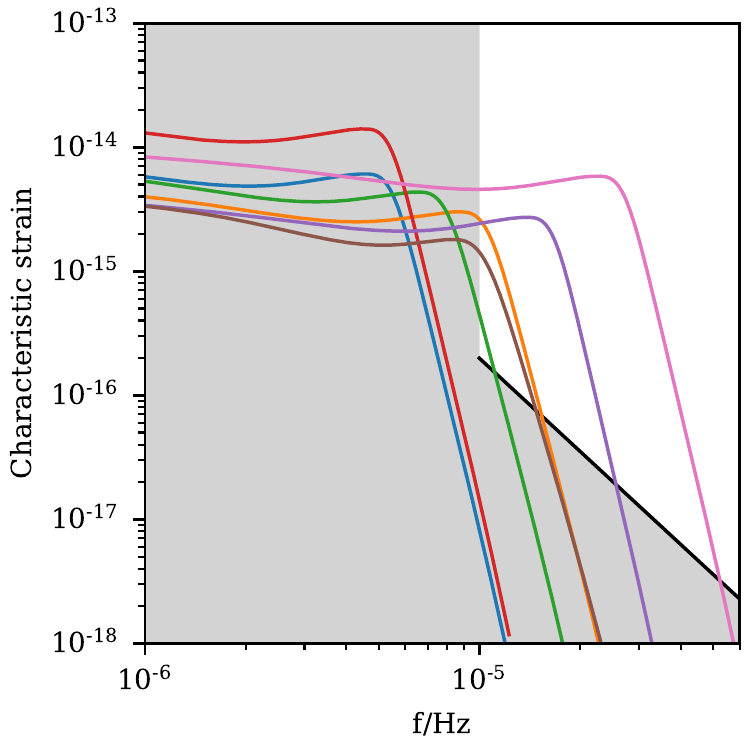}
\caption{
Left:
The evolution of $S/N$ observed with PTA as a function of time before the merger.
The line colors correspond to different binaries as in \autoref{fig:binary_params}.
The shaded region is likely to be unobservable as resolved targets.
Right:
The characteristic strain of the binaries during the final stages of their inspiral and merger.
The black line shows the planned LISA sensitivity curve.
The shaded region is unobservable with the assumed sensitivity.
}
\label{fig:gw_signals}
\end{figure*}

The right panel of \autoref{fig:gw_signals} shows the resulting characteristic
strain spectra together with the LISA sensitivity curve which is here
assumed to extend down to $f=10^{-5}\,\mathrm{Hz}$.
The $S/N$ calculation yields a high value of 410 for the \bh{J}\bh{K}-binary merger,
which would therefore be easily detectable.
The \bh{H}\bh{I} merger with $S/N\sim 55$ would likely also be detectable,
while \bh{C}\bh{D} and \bh{F}\bh{G} mergers with $S/N \sim 10$ might be \edit2{marginally} detectable
if LISA reaches the assumed low frequency sensitivity.

\section{Discussion}
\label{Sec:discuss}
\edit2{
As the simulation presented here is one of the first cosmological simulations following
in detail the small-scale evolution of SMBH binaries, it is of interest to compare the results to
earlier isolated collisionless simulations which have shaped the general understanding of SMBH binary evolution.
All the SMBH binaries that formed in this simulation were efficiently hardened
by stellar interactions without any signs of stalling due to the depletion of the 
loss-cone \citep{2001ApJ...563...34M}.
This behavior is expected for triaxial galaxies (see bottom panels of \autoref{fig:galaxy_profiles}) that form through mergers
\citep[e.g.][]{2006ApJ...642L..21B,2013ApJ...773..100K}.
The eccentric orbits of the binaries are also in general similar to the orbits
seen in earlier isolated merger simulations \citep[e.g.][]{2009ApJ...695..455B,2011ApJ...732...89K,2020MNRAS.497..739N}, which is not too surprising given that merger
simulations typically attempt to reproduce the nearly parabolic galaxy merger
orbits that are typically found in cosmological simulations \citep{2006A&A...445..403K}. 
The combined effect of eccentric binary orbits and efficient stellar hardening also result in 
SMBH merger timescales which are in good agreement with the earlier simulations.
}

\edit2{
A fundamental difference between the simulations presented here and earlier collisionless merger simulations is 
the fact that our simulations are run in a full cosmological setting and also include hydrodynamics, star formation and 
BH feedback physics. 
This allows exploring mergers with a more realistic range of parameters compared to
the more simplified systems typically used in isolated simulations.
However, the general good agreement with earlier isolated merger studies suggests
that the various idealizations used in these simulations,
such as spherical and isotropic galaxy models,
still allow for capturing the general behavior of SMBHs also seen in more realistic galaxies.
On the other hand, the lower eccentricity \bh{J}\bh{K}-binary was formed in
a galaxy merger with a wide inspiraling orbit, and is thus an example of a SMBH merger from a less explored
region of the parameter space.
}

\edit2{
A more detailed understanding of the binary evolution process could potentially be achieved by studying a wide range 
of SMBH mergers in different environments. This in turn could be used 
to improve the semi-analytic models used to model the statistics of the cosmologically merging
SMBH populations.
Although the number of binaries seen here is too small to draw any firm conclusions,
the distribution of binary eccentricities seen here does not appear to resemble the
simple distributions, e.g.\ uniform or constant, that have been used in recent studies
using such models \citep[e.g.][]{2017MNRAS.471.4508K,2019MNRAS.486.4044B,2022MNRAS.509.3488I},
which could potentially impact the validity of these results.
}

\edit2{
The details of the SMBH binary evolution presented here depend on the interactions
with the stellar component in the center of the galaxies. In cosmological simulations 
it is challenging to resolve the central regions of galaxies, as the particle masses and 
gravitational softening lengths have to be sufficiently large for the simulations to be computationally
feasible, with this challenge even increasing when higher-accuracy SMBH dynamics, such as in 
\ketju{}, is included. 
These resolution limitations appear to affect our simulation to some extent,
as the central galaxy density profiles were seen to flatten, resulting in cores on scales 
comparable to the softening length. However, the galaxies are still in general well resolved with 
moderately high numbers of a few hundred thousand stellar particles within their half-mass radii.
}

\edit2{
The artificially cored density profiles may affect the SMBH binary dynamics
by somewhat reducing the number of stellar particles interacting strongly with the binary,
which would also reduce the binary hardening rate.
The orbits of the particles are also likely to differ from those in a higher-resolution simulation, but how this would affect the dynamics is unclear.
The stellar particle counts used to resolve each galaxy,
or equivalently the stellar particle masses, may also affect the hardening rate,
as isolated $N$-body simulations have found that the binary hardening rates typically 
converge at $\sim 10^6$ stellar particles or SMBH to stellar particle mass ratios of $\sim 5000$.
Lower resolutions typically lead to higher hardening rates due
to both numerical relaxation effects and Brownian motion of the binary
\citep[e.g.][]{2013ApJ...773..100K,2016MNRAS.461.1023B}.
However, the resolution dependence of the hardening rate is rather low in triaxial systems
that form after mergers \citep[e.g.][]{2006ApJ...642L..21B,2011ApJ...732...89K,2017ApJ...840...53R,2017MNRAS.464.2301G},
and the gravitational softening used here in the stellar interactions should also further reduce the resolution
dependence caused by relaxation \citep{2017MNRAS.464.2301G}.
}

\edit2{
The similarity of the binary evolution seen here to that of
earlier isolated simulations
suggests that the potential issues described above do not compromise our
results to any significant degree.
The various convergence results found in isolated $N$-body studies
may also not generalize to full cosmological simulations, as the structure of the dynamically
formed galaxies are significantly different from the idealized models used in the 
isolated simulations.
Further work will also be needed to understand the convergence behavior of the simulations and to produce
galaxies with sufficiently resolved central regions.
In addition to simply increasing the computational effort in order to decrease the 
particle masses and softening lengths, the key effort required to achieve this will be developing more physically
motivated star formation and BH feedback models, which also account for the dynamically resolved 
BH binary phase. 
}

\edit2{
The evolution of the $M_\bullet\textnormal{--}\sigma_\star$ relation due to SMBH
recoil kick effects has already been previously studied in isolated merger simulations
\citep{2011MNRAS.412.2154B}, where the main effect of the kick was to displace the SMBHs
from the high-density galactic centers and thus resulting in a reduced gas accretion history. 
In addition, several studies of the impact of SMBH recoil kicks have been performed using 
semi-analytic techniques (e.g. \citealt{2007ApJ...663L...5V,2015MNRAS.446...38G,2020MNRAS.495.4681I}), with some studies 
finding that very strong superkicks, with velocities in excess of $\gtrsim 5000\,\mathrm{km\,s^{-1}}$ could potentially even kick the SMBHs out of the most massive 
galaxies thus producing scatter on the $M_\bullet\textnormal{--}\sigma_\star$ relation.
We showed in this study an explicit example of this mechanism,
demonstrating that kicks in excess of  $\gtrsim 2000\,\mathrm{km\,s^{-1}}$ are
in fact able to eject the SMBH from a massive galaxy in 
a dynamically resolved cosmological zoom-in simulation and thus impact the evolution of the galaxy on the $M_\bullet\textnormal{--}\sigma_\star$ plane.
}

\edit2{
For the GW signals emitted by the binaries we here evaluated only their detectability as
resolved sources, since the number of binaries in our study is far too low for
evaluating the contribution to the stochastic GW background.
Many of the binaries were found to be detectable with PTAs, which is not too surprising given that 
statistical studies have found massive binaries at $z\sim 0.5$ to be most likely the first individually 
detected sources \citep{2015MNRAS.451.2417R}.
However, it is probable that the stochastic background will be detected first \citep[e.g.][]{2018MNRAS.477..964K},
particularly given the moderately high eccentricities of the binaries seen here,
which will tend to increase the background signal if they are representative of
the typical case \citep{2017MNRAS.471.4508K}.
}

\edit2{
Some of the binaries were also found to be detectable with LISA, should they
merge during its observation period.
This is naturally quite unlikely for any individual system due to the short planned
operation time of only a few years, so this calculation mainly serves as a demonstration
that the full orbital evolution of targets relevant for LISA can be followed in a
cosmological setting using \ketju{}.
Lower-mass SMBH systems which will be more readily detectable with LISA are typically found in
late-type gas-rich galaxies, and thus the simulations require improved accretion and star formation models in order 
to be modeled properly with \ketju{}.
}

\vfill\eject 
\section{Conclusions}
\label{Sec:conclusions}

We have demonstrated that the \ketju{} code can be used to model the small-scale 
dynamical evolution of dozens of SMBHs in a complex cosmological environment over long
periods of time. Seven SMBH binary systems formed during the simulation, and they were all driven
to merger by stellar interactions without any signs of stalling.
The detailed 
dynamical evolution of the SMBH binaries naturally depends on how well the central stellar component is resolved,
\edit2{which is the main limitation of our present study}. 
However, it is encouraging that the results obtained here agree well with simulations
of similar systems run in isolation at higher spatial \edit2{and mass resolutions}.

Our simulated SMBH binaries typically formed on quite eccentric orbits,
which \edit2{highlights the need to include the effects of eccentricity} for
correctly capturing the binary evolution in a majority of the cases. In addition, modeling systems 
of multiple interacting SMBHs is important for correctly capturing the SMBH merger timescales, as
such situations will naturally occur in a cosmological setting.  
The displacement and ejection of SMBHs \edit2{by GW recoil kicks} will introduce scatter to
the observed $M_\bullet \textnormal{--} \sigma_\star$ relation, but as demonstrated here,
it is also possible for a galaxy to recover and 
evolve back onto the relation under the right circumstances. 

Finally, we showed that many of the simulated SMBHs in this study would be potentially detectable with PTAs and LISA, if they had existed in the real Universe 
at the observed redshifts.
In particular PTAs could potentially detect GW signals from the eccentric orbital
phase $\sim 10 \, \mathrm{Myr}$ before the merger, whereas LISA
would only be able to detect the final stages of the SMBH mergers simulated in this study,
given their relatively large masses. \edit2{Simultaneous modeling of the accurate small-scale dynamics and gas dynamics will 
be in particular important for making GW predictions for LISA, as it will be mostly sensitive to SMBHs in the mass range of $M_\bullet \sim 10^{5} \textnormal{--} 10^{7} M_{\odot}$, which are
expected to reside in late-type gas-rich galaxies.}

\begin{acknowledgments} 
\modulolinenumbers[100] 

M.M., P.H.J., S.L., and A.R. acknowledge the support
by the European Research Council via ERC Consolidator Grant KETJU (no. 818930) and the support of the Academy of Finland grant 339127. 

T.N. acknowledges support from the Deutsche Forschungsgemeinschaft 
(DFG, German Research Foundation) under Germany's Excellence Strategy - EXC-2094 - 390783311 from the DFG Cluster of Excellence ``ORIGINS''.

The numerical simulations used computational resources provided by 
the CSC -- IT Center for Science, Finland.
\end{acknowledgments}

\software{
\ketju{} \citep{2017ApJ...840...53R,2020MNRAS.492.4131R},
\gadget{} \citep{2005MNRAS.364.1105S},
NumPy \citep{numpy},
SciPy \citep{scipy},
Matplotlib \citep{matplotlib},
pygad \citep{2020MNRAS.496..152R},
MUSIC \citep{2011MNRAS.415.2101H,2013ascl.soft11011H},
BOWIE \citep{2019MNRAS.483.3108K}.
}

\bibliographystyle{aasjournal}
\bibliography{refs}

\begin{thebibliography}{}
\expandafter\ifx\csname natexlab\endcsname\relax\def\natexlab#1{#1}\fi
\providecommand{\url}[1]{\href{#1}{#1}}
\providecommand{\dodoi}[1]{doi:~\href{http://doi.org/#1}{\nolinkurl{#1}}}
\providecommand{\doeprint}[1]{\href{http://ascl.net/#1}{\nolinkurl{http://ascl.net/#1}}}
\providecommand{\doarXiv}[1]{\href{https://arxiv.org/abs/#1}{\nolinkurl{https://arxiv.org/abs/#1}}}

\bibitem[{{Alam} {et~al.}(2021){Alam}, {Arzoumanian}, {Baker}, {Blumer},
  {Bohler}, {Brazier}, {Brook}, {Burke-Spolaor}, {Caballero}, {Camuccio},
  {Chamberlain}, {Chatterjee}, {Cordes}, {Cornish}, {Crawford}, {Cromartie},
  {Decesar}, {Demorest}, {Dolch}, {Ellis}, {Ferdman}, {Ferrara}, {Fiore},
  {Fonseca}, {Garcia}, {Garver-Daniels}, {Gentile}, {Good}, {Gusdorff},
  {Halmrast}, {Hazboun}, {Islo}, {Jennings}, {Jessup}, {Jones}, {Kaiser},
  {Kaplan}, {Kelley}, {Key}, {Lam}, {Lazio}, {Lorimer}, {Luo}, {Lynch},
  {Madison}, {Maraccini}, {McLaughlin}, {Mingarelli}, {Ng}, {Nguyen}, {Nice},
  {Pennucci}, {Pol}, {Ramette}, {Ransom}, {Ray}, {Shapiro-Albert}, {Siemens},
  {Simon}, {Spiewak}, {Stairs}, {Stinebring}, {Stovall}, {Swiggum}, {Taylor},
  {Tripepi}, {Vallisneri}, {Vigeland}, {Witt}, {Zhu}, \& {Nanograv
  Collaboration}}]{2021ApJS..252....5A}
{Alam}, M.~F., {Arzoumanian}, Z., {Baker}, P.~T., {et~al.} 2021, \apjs, 252, 5,
  \dodoi{10.3847/1538-4365/abc6a1}

\bibitem[{{Amaro-Seoane} {et~al.}(2017){Amaro-Seoane}, {Audley}, {Babak},
  {Baker}, {Barausse}, {Bender}, {Berti}, {Binetruy}, {Born}, {Bortoluzzi},
  {Camp}, {Caprini}, {Cardoso}, {Colpi}, {Conklin}, {Cornish}, {Cutler},
  {Danzmann}, {Dolesi}, {Ferraioli}, {Ferroni}, {Fitzsimons}, {Gair}, {Gesa
  Bote}, {Giardini}, {Gibert}, {Grimani}, {Halloin}, {Heinzel}, {Hertog},
  {Hewitson}, {Holley-Bockelmann}, {Hollington}, {Hueller}, {Inchauspe},
  {Jetzer}, {Karnesis}, {Killow}, {Klein}, {Klipstein}, {Korsakova}, {Larson},
  {Livas}, {Lloro}, {Man}, {Mance}, {Martino}, {Mateos}, {McKenzie},
  {McWilliams}, {Miller}, {Mueller}, {Nardini}, {Nelemans}, {Nofrarias},
  {Petiteau}, {Pivato}, {Plagnol}, {Porter}, {Reiche}, {Robertson},
  {Robertson}, {Rossi}, {Russano}, {Schutz}, {Sesana}, {Shoemaker}, {Slutsky},
  {Sopuerta}, {Sumner}, {Tamanini}, {Thorpe}, {Troebs}, {Vallisneri},
  {Vecchio}, {Vetrugno}, {Vitale}, {Volonteri}, {Wanner}, {Ward}, {Wass},
  {Weber}, {Ziemer}, \& {Zweifel}}]{2017arXiv170200786A}
{Amaro-Seoane}, P., {Audley}, H., {Babak}, S., {et~al.} 2017, arXiv e-prints,
  arXiv:1702.00786.
\newblock \doarXiv{1702.00786}

\bibitem[{{Arzoumanian} {et~al.}(2020){Arzoumanian}, {Baker}, {Blumer},
  {B{\'e}csy}, {Brazier}, {Brook}, {Burke-Spolaor}, {Chatterjee}, {Chen},
  {Cordes}, {Cornish}, {Crawford}, {Cromartie}, {Decesar}, {Demorest}, {Dolch},
  {Ellis}, {Ferrara}, {Fiore}, {Fonseca}, {Garver-Daniels}, {Gentile}, {Good},
  {Hazboun}, {Holgado}, {Islo}, {Jennings}, {Jones}, {Kaiser}, {Kaplan},
  {Kelley}, {Key}, {Laal}, {Lam}, {Lazio}, {Lorimer}, {Luo}, {Lynch},
  {Madison}, {McLaughlin}, {Mingarelli}, {Ng}, {Nice}, {Pennucci}, {Pol},
  {Ransom}, {Ray}, {Shapiro-Albert}, {Siemens}, {Simon}, {Spiewak}, {Stairs},
  {Stinebring}, {Stovall}, {Sun}, {Swiggum}, {Taylor}, {Turner}, {Vallisneri},
  {Vigeland}, {Witt}, \& {Nanograv Collaboration}}]{2020ApJ...905L..34A}
{Arzoumanian}, Z., {Baker}, P.~T., {Blumer}, H., {et~al.} 2020, \apjl, 905,
  L34, \dodoi{10.3847/2041-8213/abd401}

\bibitem[{{Aumer} {et~al.}(2013){Aumer}, {White}, {Naab}, \&
  {Scannapieco}}]{2013MNRAS.434.3142A}
{Aumer}, M., {White}, S. D.~M., {Naab}, T., \& {Scannapieco}, C. 2013, \mnras,
  434, 3142, \dodoi{10.1093/mnras/stt1230}

\bibitem[{{Begelman} {et~al.}(1980){Begelman}, {Blandford}, \&
  {Rees}}]{begelman1980}
{Begelman}, M.~C., {Blandford}, R.~D., \& {Rees}, M.~J. 1980, \nat, 287, 307,
  \dodoi{10.1038/287307a0}

\bibitem[{{Berczik} {et~al.}(2006){Berczik}, {Merritt}, {Spurzem}, \&
  {Bischof}}]{2006ApJ...642L..21B}
{Berczik}, P., {Merritt}, D., {Spurzem}, R., \& {Bischof}, H.-P. 2006, \apjl,
  642, L21, \dodoi{10.1086/504426}

\bibitem[{{Berentzen} {et~al.}(2009){Berentzen}, {Preto}, {Berczik}, {Merritt},
  \& {Spurzem}}]{2009ApJ...695..455B}
{Berentzen}, I., {Preto}, M., {Berczik}, P., {Merritt}, D., \& {Spurzem}, R.
  2009, \apj, 695, 455, \dodoi{10.1088/0004-637X/695/1/455}

\bibitem[{{Bezanson} {et~al.}(2018){Bezanson}, {van der Wel}, {Pacifici},
  {Noeske}, {Bari{\v{s}}i{\'c}}, {Bell}, {Brammer}, {Calhau}, {Chauke}, {van
  Dokkum}, {Franx}, {Gallazzi}, {van Houdt}, {Labb{\'e}}, {Maseda},
  {Mu{\~n}os-Mateos}, {Muzzin}, {van de Sande}, {Sobral}, {Straatman}, \&
  {Wu}}]{2018ApJ...858...60B}
{Bezanson}, R., {van der Wel}, A., {Pacifici}, C., {et~al.} 2018, \apj, 858,
  60, \dodoi{10.3847/1538-4357/aabc55}

\bibitem[{{Blanchet}(2014)}]{2014LRR....17....2B}
{Blanchet}, L. 2014, Living Reviews in Relativity, 17, 2,
  \dodoi{10.12942/lrr-2014-2}

\bibitem[{{Blecha} {et~al.}(2011){Blecha}, {Cox}, {Loeb}, \&
  {Hernquist}}]{2011MNRAS.412.2154B}
{Blecha}, L., {Cox}, T.~J., {Loeb}, A., \& {Hernquist}, L. 2011, \mnras, 412,
  2154, \dodoi{10.1111/j.1365-2966.2010.18042.x}

\bibitem[{{Bonetti} {et~al.}(2019){Bonetti}, {Sesana}, {Haardt}, {Barausse}, \&
  {Colpi}}]{2019MNRAS.486.4044B}
{Bonetti}, M., {Sesana}, A., {Haardt}, F., {Barausse}, E., \& {Colpi}, M. 2019,
  \mnras, 486, 4044, \dodoi{10.1093/mnras/stz903}

\bibitem[{{Bortolas} {et~al.}(2016){Bortolas}, {Gualandris}, {Dotti}, {Spera},
  \& {Mapelli}}]{2016MNRAS.461.1023B}
{Bortolas}, E., {Gualandris}, A., {Dotti}, M., {Spera}, M., \& {Mapelli}, M.
  2016, \mnras, 461, 1023, \dodoi{10.1093/mnras/stw1372}

\bibitem[{{Campanelli} {et~al.}(2007){Campanelli}, {Lousto}, {Zlochower}, \&
  {Merritt}}]{2007ApJ...659L...5C}
{Campanelli}, M., {Lousto}, C., {Zlochower}, Y., \& {Merritt}, D. 2007, \apjl,
  659, L5, \dodoi{10.1086/516712}

\bibitem[{{Comerford} {et~al.}(2015){Comerford}, {Pooley}, {Barrows}, {Greene},
  {Zakamska}, {Madejski}, \& {Cooper}}]{2015ApJ...806..219C}
{Comerford}, J.~M., {Pooley}, D., {Barrows}, R.~S., {et~al.} 2015, \apj, 806,
  219, \dodoi{10.1088/0004-637X/806/2/219}

\bibitem[{{Eisenreich} {et~al.}(2017){Eisenreich}, {Naab}, {Choi}, {Ostriker},
  \& {Emsellem}}]{2017MNRAS.468..751E}
{Eisenreich}, M., {Naab}, T., {Choi}, E., {Ostriker}, J.~P., \& {Emsellem}, E.
  2017, \mnras, 468, 751, \dodoi{10.1093/mnras/stx473}

\bibitem[{{Gerosa} \& {Sesana}(2015)}]{2015MNRAS.446...38G}
{Gerosa}, D., \& {Sesana}, A. 2015, \mnras, 446, 38,
  \dodoi{10.1093/mnras/stu2049}

\bibitem[{{Gualandris} {et~al.}(2017){Gualandris}, {Read}, {Dehnen}, \&
  {Bortolas}}]{2017MNRAS.464.2301G}
{Gualandris}, A., {Read}, J.~I., {Dehnen}, W., \& {Bortolas}, E. 2017, \mnras,
  464, 2301, \dodoi{10.1093/mnras/stw2528}

\bibitem[{{Hahn} \& {Abel}(2011)}]{2011MNRAS.415.2101H}
{Hahn}, O., \& {Abel}, T. 2011, \mnras, 415, 2101,
  \dodoi{10.1111/j.1365-2966.2011.18820.x}

\bibitem[{{Hahn} \& {Abel}(2013)}]{2013ascl.soft11011H}
{Hahn}, O., \& {Abel}, T. 2013, {MUSIC: MUlti-Scale Initial Conditions}.
\newblock \doeprint{1311.011}

\bibitem[{{Harris} {et~al.}(2020){Harris}, {Millman}, {van der Walt},
  {Gommers}, {Virtanen}, {Cournapeau}, {Wieser}, {Taylor}, {Berg}, {Smith},
  {Kern}, {Picus}, {Hoyer}, {van Kerkwijk}, {Brett}, {Haldane}, {del R{\'\i}o},
  {Wiebe}, {Peterson}, {G{\'e}rard-Marchant}, {Sheppard}, {Reddy}, {Weckesser},
  {Abbasi}, {Gohlke}, \& {Oliphant}}]{numpy}
{Harris}, C.~R., {Millman}, K.~J., {van der Walt}, S.~J., {et~al.} 2020, \nat,
  585, 357, \dodoi{10.1038/s41586-020-2649-2}

\bibitem[{{Hills} \& {Fullerton}(1980)}]{1980AJ.....85.1281H}
{Hills}, J.~G., \& {Fullerton}, L.~W. 1980, \aj, 85, 1281,
  \dodoi{10.1086/112798}

\bibitem[{{Hu} {et~al.}(2014){Hu}, {Naab}, {Walch}, {Moster}, \&
  {Oser}}]{2014MNRAS.443.1173H}
{Hu}, C.-Y., {Naab}, T., {Walch}, S., {Moster}, B.~P., \& {Oser}, L. 2014,
  \mnras, 443, 1173, \dodoi{10.1093/mnras/stu1187}

\bibitem[{{Huerta} {et~al.}(2015){Huerta}, {McWilliams}, {Gair}, \&
  {Taylor}}]{2015PhRvD..92f3010H}
{Huerta}, E.~A., {McWilliams}, S.~T., {Gair}, J.~R., \& {Taylor}, S.~R. 2015,
  \prd, 92, 063010, \dodoi{10.1103/PhysRevD.92.063010}

\bibitem[{{Hunter}(2007)}]{matplotlib}
{Hunter}, J.~D. 2007, Computing in Science and Engineering, 9, 90,
  \dodoi{10.1109/MCSE.2007.55}

\bibitem[{{Husa} {et~al.}(2016){Husa}, {Khan}, {Hannam}, {P{\"u}rrer}, {Ohme},
  {Forteza}, \& {Boh{\'e}}}]{2016PhRvD..93d4006H}
{Husa}, S., {Khan}, S., {Hannam}, M., {et~al.} 2016, \prd, 93, 044006,
  \dodoi{10.1103/PhysRevD.93.044006}

\bibitem[{{Izquierdo-Villalba} {et~al.}(2020){Izquierdo-Villalba}, {Bonoli},
  {Dotti}, {Sesana}, {Rosas-Guevara}, \& {Spinoso}}]{2020MNRAS.495.4681I}
{Izquierdo-Villalba}, D., {Bonoli}, S., {Dotti}, M., {et~al.} 2020, \mnras,
  495, 4681, \dodoi{10.1093/mnras/staa1399}

\bibitem[{{Izquierdo-Villalba} {et~al.}(2022){Izquierdo-Villalba}, {Sesana},
  {Bonoli}, \& {Colpi}}]{2022MNRAS.509.3488I}
{Izquierdo-Villalba}, D., {Sesana}, A., {Bonoli}, S., \& {Colpi}, M. 2022,
  \mnras, 509, 3488, \dodoi{10.1093/mnras/stab3239}

\bibitem[{{Johansson} {et~al.}(2009{\natexlab{a}}){Johansson}, {Burkert}, \&
  {Naab}}]{2009ApJ...707L.184J}
{Johansson}, P.~H., {Burkert}, A., \& {Naab}, T. 2009{\natexlab{a}}, \apjl,
  707, L184, \dodoi{10.1088/0004-637X/707/2/L184}

\bibitem[{{Johansson} {et~al.}(2009{\natexlab{b}}){Johansson}, {Naab}, \&
  {Burkert}}]{2009ApJ...690..802J}
{Johansson}, P.~H., {Naab}, T., \& {Burkert}, A. 2009{\natexlab{b}}, \apj, 690,
  802, \dodoi{10.1088/0004-637X/690/1/802}

\bibitem[{{Johansson} {et~al.}(2012){Johansson}, {Naab}, \&
  {Ostriker}}]{2012ApJ...754..115J}
{Johansson}, P.~H., {Naab}, T., \& {Ostriker}, J.~P. 2012, \apj, 754, 115,
  \dodoi{10.1088/0004-637X/754/2/115}

\bibitem[{{Katz} \& {Larson}(2019)}]{2019MNRAS.483.3108K}
{Katz}, M.~L., \& {Larson}, S.~L. 2019, \mnras, 483, 3108,
  \dodoi{10.1093/mnras/sty3321}

\bibitem[{{Kelley} {et~al.}(2017){Kelley}, {Blecha}, {Hernquist}, {Sesana}, \&
  {Taylor}}]{2017MNRAS.471.4508K}
{Kelley}, L.~Z., {Blecha}, L., {Hernquist}, L., {Sesana}, A., \& {Taylor},
  S.~R. 2017, \mnras, 471, 4508, \dodoi{10.1093/mnras/stx1638}

\bibitem[{{Kelley} {et~al.}(2018){Kelley}, {Blecha}, {Hernquist}, {Sesana}, \&
  {Taylor}}]{2018MNRAS.477..964K}
{Kelley}, L.~Z., {Blecha}, L., {Hernquist}, L., {Sesana}, A., \& {Taylor},
  S.~R. 2018, \mnras, 477, 964, \dodoi{10.1093/mnras/sty689}

\bibitem[{{Khan} {et~al.}(2016{\natexlab{a}}){Khan}, {Fiacconi}, {Mayer},
  {Berczik}, \& {Just}}]{2016ApJ...828...73K}
{Khan}, F.~M., {Fiacconi}, D., {Mayer}, L., {Berczik}, P., \& {Just}, A.
  2016{\natexlab{a}}, \apj, 828, 73, \dodoi{10.3847/0004-637X/828/2/73}

\bibitem[{{Khan} {et~al.}(2013){Khan}, {Holley-Bockelmann}, {Berczik}, \&
  {Just}}]{2013ApJ...773..100K}
{Khan}, F.~M., {Holley-Bockelmann}, K., {Berczik}, P., \& {Just}, A. 2013,
  \apj, 773, 100, \dodoi{10.1088/0004-637X/773/2/100}

\bibitem[{{Khan} {et~al.}(2011){Khan}, {Just}, \&
  {Merritt}}]{2011ApJ...732...89K}
{Khan}, F.~M., {Just}, A., \& {Merritt}, D. 2011, \apj, 732, 89,
  \dodoi{10.1088/0004-637X/732/2/89}

\bibitem[{{Khan} {et~al.}(2016{\natexlab{b}}){Khan}, {Husa}, {Hannam}, {Ohme},
  {P{\"u}rrer}, {Forteza}, \& {Boh{\'e}}}]{2016PhRvD..93d4007K}
{Khan}, S., {Husa}, S., {Hannam}, M., {et~al.} 2016{\natexlab{b}}, \prd, 93,
  044007, \dodoi{10.1103/PhysRevD.93.044007}

\bibitem[{{Khochfar} \& {Burkert}(2006)}]{2006A&A...445..403K}
{Khochfar}, S., \& {Burkert}, A. 2006, \aap, 445, 403,
  \dodoi{10.1051/0004-6361:20053241}

\bibitem[{{Kormendy} \& {Ho}(2013)}]{kormendy2013}
{Kormendy}, J., \& {Ho}, L.~C. 2013, \araa, 51, 511,
  \dodoi{10.1146/annurev-astro-082708-101811}

\bibitem[{{Lidov}(1962)}]{1962P&SS....9..719L}
{Lidov}, M.~L. 1962, \planss, 9, 719, \dodoi{10.1016/0032-0633(62)90129-0}

\bibitem[{{Lim} \& {Rodriguez}(2020)}]{2020PhRvD.102f4033L}
{Lim}, H., \& {Rodriguez}, C.~L. 2020, \prd, 102, 064033,
  \dodoi{10.1103/PhysRevD.102.064033}

\bibitem[{{Lousto} {et~al.}(2010){Lousto}, {Nakano}, {Zlochower}, \&
  {Campanelli}}]{2010PhRvD..81h4023L}
{Lousto}, C.~O., {Nakano}, H., {Zlochower}, Y., \& {Campanelli}, M. 2010, \prd,
  81, 084023, \dodoi{10.1103/PhysRevD.81.084023}

\bibitem[{{Mannerkoski} {et~al.}(2019){Mannerkoski}, {Johansson}, {Pihajoki},
  {Rantala}, \& {Naab}}]{2019ApJ...887...35M}
{Mannerkoski}, M., {Johansson}, P.~H., {Pihajoki}, P., {Rantala}, A., \&
  {Naab}, T. 2019, \apj, 887, 35, \dodoi{10.3847/1538-4357/ab52f9}

\bibitem[{{Mannerkoski} {et~al.}(2021){Mannerkoski}, {Johansson}, {Rantala},
  {Naab}, \& {Liao}}]{2021ApJ...912L..20M}
{Mannerkoski}, M., {Johansson}, P.~H., {Rantala}, A., {Naab}, T., \& {Liao}, S.
  2021, \apjl, 912, L20, \dodoi{10.3847/2041-8213/abf9a5}

\bibitem[{{Memmesheimer} {et~al.}(2004){Memmesheimer}, {Gopakumar}, \&
  {Sch{\"a}fer}}]{2004PhRvD..70j4011M}
{Memmesheimer}, R.-M., {Gopakumar}, A., \& {Sch{\"a}fer}, G. 2004, \prd, 70,
  104011, \dodoi{10.1103/PhysRevD.70.104011}

\bibitem[{{Milosavljevi{\'c}} \& {Merritt}(2001)}]{2001ApJ...563...34M}
{Milosavljevi{\'c}}, M., \& {Merritt}, D. 2001, \apj, 563, 34,
  \dodoi{10.1086/323830}

\bibitem[{{Naab} \& {Ostriker}(2017)}]{2017ARA&A..55...59N}
{Naab}, T., \& {Ostriker}, J.~P. 2017, \araa, 55, 59,
  \dodoi{10.1146/annurev-astro-081913-040019}

\bibitem[{{Nasim} {et~al.}(2020){Nasim}, {Gualandris}, {Read}, {Dehnen},
  {Delorme}, \& {Antonini}}]{2020MNRAS.497..739N}
{Nasim}, I., {Gualandris}, A., {Read}, J., {et~al.} 2020, \mnras, 497, 739,
  \dodoi{10.1093/mnras/staa1896}

\bibitem[{{Nasim} {et~al.}(2021){Nasim}, {Gualandris}, {Read}, {Antonini},
  {Dehnen}, \& {Delorme}}]{2021MNRAS.502.4794N}
{Nasim}, I.~T., {Gualandris}, A., {Read}, J.~I., {et~al.} 2021, \mnras, 502,
  4794, \dodoi{10.1093/mnras/stab435}

\bibitem[{{Peters}(1964)}]{1964PhRv..136.1224P}
{Peters}, P.~C. 1964, Physical Review, 136, 1224,
  \dodoi{10.1103/PhysRev.136.B1224}

\bibitem[{{Planck Collaboration} {et~al.}(2020){Planck Collaboration},
  {Aghanim}, {Akrami}, {Ashdown}, {Aumont}, {Baccigalupi}, {Ballardini},
  {Banday}, {Barreiro}, {Bartolo}, {Basak}, {Battye}, {Benabed}, {Bernard},
  {Bersanelli}, {Bielewicz}, {Bock}, {Bond}, {Borrill}, {Bouchet}, {Boulanger},
  {Bucher}, {Burigana}, {Butler}, {Calabrese}, {Cardoso}, {Carron},
  {Challinor}, {Chiang}, {Chluba}, {Colombo}, {Combet}, {Contreras}, {Crill},
  {Cuttaia}, {de Bernardis}, {de Zotti}, {Delabrouille}, {Delouis}, {Di
  Valentino}, {Diego}, {Dor{\'e}}, {Douspis}, {Ducout}, {Dupac}, {Dusini},
  {Efstathiou}, {Elsner}, {En{\ss}lin}, {Eriksen}, {Fantaye}, {Farhang},
  {Fergusson}, {Fernandez-Cobos}, {Finelli}, {Forastieri}, {Frailis},
  {Fraisse}, {Franceschi}, {Frolov}, {Galeotta}, {Galli}, {Ganga},
  {G{\'e}nova-Santos}, {Gerbino}, {Ghosh}, {Gonz{\'a}lez-Nuevo}, {G{\'o}rski},
  {Gratton}, {Gruppuso}, {Gudmundsson}, {Hamann}, {Handley}, {Hansen},
  {Herranz}, {Hildebrandt}, {Hivon}, {Huang}, {Jaffe}, {Jones}, {Karakci},
  {Keih{\"a}nen}, {Keskitalo}, {Kiiveri}, {Kim}, {Kisner}, {Knox},
  {Krachmalnicoff}, {Kunz}, {Kurki-Suonio}, {Lagache}, {Lamarre}, {Lasenby},
  {Lattanzi}, {Lawrence}, {Le Jeune}, {Lemos}, {Lesgourgues}, {Levrier},
  {Lewis}, {Liguori}, {Lilje}, {Lilley}, {Lindholm}, {L{\'o}pez-Caniego},
  {Lubin}, {Ma}, {Mac{\'\i}as-P{\'e}rez}, {Maggio}, {Maino}, {Mandolesi},
  {Mangilli}, {Marcos-Caballero}, {Maris}, {Martin}, {Martinelli},
  {Mart{\'\i}nez-Gonz{\'a}lez}, {Matarrese}, {Mauri}, {McEwen}, {Meinhold},
  {Melchiorri}, {Mennella}, {Migliaccio}, {Millea}, {Mitra},
  {Miville-Desch{\^e}nes}, {Molinari}, {Montier}, {Morgante}, {Moss}, {Natoli},
  {N{\o}rgaard-Nielsen}, {Pagano}, {Paoletti}, {Partridge}, {Patanchon},
  {Peiris}, {Perrotta}, {Pettorino}, {Piacentini}, {Polastri}, {Polenta},
  {Puget}, {Rachen}, {Reinecke}, {Remazeilles}, {Renzi}, {Rocha}, {Rosset},
  {Roudier}, {Rubi{\~n}o-Mart{\'\i}n}, {Ruiz-Granados}, {Salvati}, {Sandri},
  {Savelainen}, {Scott}, {Shellard}, {Sirignano}, {Sirri}, {Spencer},
  {Sunyaev}, {Suur-Uski}, {Tauber}, {Tavagnacco}, {Tenti}, {Toffolatti},
  {Tomasi}, {Trombetti}, {Valenziano}, {Valiviita}, {Van Tent}, {Vibert},
  {Vielva}, {Villa}, {Vittorio}, {Wandelt}, {Wehus}, {White}, {White},
  {Zacchei}, \& {Zonca}}]{2020A&A...641A...6P}
{Planck Collaboration}, {Aghanim}, N., {Akrami}, Y., {et~al.} 2020, \aap, 641,
  A6, \dodoi{10.1051/0004-6361/201833910}

\bibitem[{{Quinlan}(1996)}]{1996NewA....1...35Q}
{Quinlan}, G.~D. 1996, \na, 1, 35, \dodoi{10.1016/S1384-1076(96)00003-6}

\bibitem[{{Rantala} {et~al.}(2018){Rantala}, {Johansson}, {Naab}, {Thomas}, \&
  {Frigo}}]{2018ApJ...864..113R}
{Rantala}, A., {Johansson}, P.~H., {Naab}, T., {Thomas}, J., \& {Frigo}, M.
  2018, \apj, 864, 113, \dodoi{10.3847/1538-4357/aada47}

\bibitem[{{Rantala} {et~al.}(2017){Rantala}, {Pihajoki}, {Johansson}, {Naab},
  {Lah{\'e}n}, \& {Sawala}}]{2017ApJ...840...53R}
{Rantala}, A., {Pihajoki}, P., {Johansson}, P.~H., {et~al.} 2017, \apj, 840,
  53, \dodoi{10.3847/1538-4357/aa6d65}

\bibitem[{{Rantala} {et~al.}(2020){Rantala}, {Pihajoki}, {Mannerkoski},
  {Johansson}, \& {Naab}}]{2020MNRAS.492.4131R}
{Rantala}, A., {Pihajoki}, P., {Mannerkoski}, M., {Johansson}, P.~H., \&
  {Naab}, T. 2020, \mnras, 492, 4131, \dodoi{10.1093/mnras/staa084}

\bibitem[{{Rauch} \& {Tremaine}(1996)}]{1996NewA....1..149R}
{Rauch}, K.~P., \& {Tremaine}, S. 1996, \na, 1, 149,
  \dodoi{10.1016/S1384-1076(96)00012-7}

\bibitem[{{Reynolds}(2019)}]{2019NatAs...3...41R}
{Reynolds}, C.~S. 2019, Nature Astronomy, 3, 41,
  \dodoi{10.1038/s41550-018-0665-z}

\bibitem[{{Rosado} {et~al.}(2015){Rosado}, {Sesana}, \&
  {Gair}}]{2015MNRAS.451.2417R}
{Rosado}, P.~A., {Sesana}, A., \& {Gair}, J. 2015, \mnras, 451, 2417,
  \dodoi{10.1093/mnras/stv1098}

\bibitem[{{R{\"o}ttgers} {et~al.}(2020){R{\"o}ttgers}, {Naab}, {Cernetic},
  {Dav{\'e}}, {Kauffmann}, {Borthakur}, \& {Foidl}}]{2020MNRAS.496..152R}
{R{\"o}ttgers}, B., {Naab}, T., {Cernetic}, M., {et~al.} 2020, \mnras, 496,
  152, \dodoi{10.1093/mnras/staa1490}

\bibitem[{{Scannapieco} {et~al.}(2005){Scannapieco}, {Tissera}, {White}, \&
  {Springel}}]{2005MNRAS.364..552S}
{Scannapieco}, C., {Tissera}, P.~B., {White}, S.~D.~M., \& {Springel}, V. 2005,
  \mnras, 364, 552, \dodoi{10.1111/j.1365-2966.2005.09574.x}

\bibitem[{{Scannapieco} {et~al.}(2006){Scannapieco}, {Tissera}, {White}, \&
  {Springel}}]{2006MNRAS.371.1125S}
{Scannapieco}, C., {Tissera}, P.~B., {White}, S.~D.~M., \& {Springel}, V. 2006,
  \mnras, 371, 1125, \dodoi{10.1111/j.1365-2966.2006.10785.x}

\bibitem[{{Sijacki} {et~al.}(2007){Sijacki}, {Springel}, {Di Matteo}, \&
  {Hernquist}}]{2007MNRAS.380..877S}
{Sijacki}, D., {Springel}, V., {Di Matteo}, T., \& {Hernquist}, L. 2007,
  \mnras, 380, 877, \dodoi{10.1111/j.1365-2966.2007.12153.x}

\bibitem[{{Springel}(2005)}]{2005MNRAS.364.1105S}
{Springel}, V. 2005, \mnras, 364, 1105,
  \dodoi{10.1111/j.1365-2966.2005.09655.x}

\bibitem[{{Springel} {et~al.}(2005){Springel}, {Di Matteo}, \&
  {Hernquist}}]{2005MNRAS.361..776S}
{Springel}, V., {Di Matteo}, T., \& {Hernquist}, L. 2005, \mnras, 361, 776,
  \dodoi{10.1111/j.1365-2966.2005.09238.x}

\bibitem[{{Taylor} {et~al.}(2016){Taylor}, {Huerta}, {Gair}, \&
  {McWilliams}}]{2016ApJ...817...70T}
{Taylor}, S.~R., {Huerta}, E.~A., {Gair}, J.~R., \& {McWilliams}, S.~T. 2016,
  \apj, 817, 70, \dodoi{10.3847/0004-637X/817/1/70}

\bibitem[{{Thorne} \& {Hartle}(1985)}]{1985PhRvD..31.1815T}
{Thorne}, K.~S., \& {Hartle}, J.~B. 1985, \prd, 31, 1815,
  \dodoi{10.1103/PhysRevD.31.1815}

\bibitem[{{Virtanen} {et~al.}(2020){Virtanen}, {Gommers}, {Oliphant},
  {Haberland}, {Reddy}, {Cournapeau}, {Burovski}, {Peterson}, {Weckesser},
  {Bright}, {van der Walt}, {Brett}, {Wilson}, {Millman}, {Mayorov}, {Nelson},
  {Jones}, {Kern}, {Larson}, {Carey}, {Polat}, {Feng}, {Moore}, {VanderPlas},
  {Laxalde}, {Perktold}, {Cimrman}, {Henriksen}, {Quintero}, {Harris},
  {Archibald}, {Ribeiro}, {Pedregosa}, {van Mulbregt}, \& {SciPy 1. 0
  Contributors}}]{scipy}
{Virtanen}, P., {Gommers}, R., {Oliphant}, T.~E., {et~al.} 2020, Nature
  Methods, 17, 261, \dodoi{10.1038/s41592-019-0686-2}

\bibitem[{{Volonteri}(2007)}]{2007ApJ...663L...5V}
{Volonteri}, M. 2007, \apjl, 663, L5, \dodoi{10.1086/519525}

\bibitem[{{Will}(2014)}]{2014PhRvD..89d4043W}
{Will}, C.~M. 2014, \prd, 89, 044043, \dodoi{10.1103/PhysRevD.89.044043}

\bibitem[{{Zemp} {et~al.}(2011){Zemp}, {Gnedin}, {Gnedin}, \&
  {Kravtsov}}]{2011ApJS..197...30Z}
{Zemp}, M., {Gnedin}, O.~Y., {Gnedin}, N.~Y., \& {Kravtsov}, A.~V. 2011, \apjs,
  197, 30, \dodoi{10.1088/0067-0049/197/2/30}

\bibitem[{{Zlochower} \& {Lousto}(2015)}]{2015PhRvD..92b4022Z}
{Zlochower}, Y., \& {Lousto}, C.~O. 2015, \prd, 92, 024022,
  \dodoi{10.1103/PhysRevD.92.024022}

\end{thebibliography}

\end{document}